\documentclass[fleqn,twoside]{article}
\usepackage[headings]{espcrc2}

\readRCS
$Id: espcrc2.tex,v 1.2 2004/02/24 11:22:11 spepping Exp $
\usepackage{graphicx}
\usepackage[figuresright]{rotating}

\newcommand{\AmS}{{\protect\the\textfont2
  A\kern-.1667em\lower.5ex\hbox{M}\kern-.125emS}}

\newcommand*{\CC}{C${}^{{}_{++}}$}
\newcommand{\xloops}{{\sf xloops }}

\newcommand{\ginac}{{\sf  GiNaC }}
\newcommand{\ginacnosp}{{\sf  GiNaC}}

\title{\ginac - Symbolic computation with \CC}

\author{J. Vollinga\\[2ex]{Institut f\"ur Physik, Universit\"at Mainz, Germany}}

\runtitle{\ginac - Symbolic computation with \CC}
\runauthor{J. Vollinga}

\begin{document}

\begin{abstract}
We give an introduction to the \CC\ library \ginacnosp, which 
extends the \CC\ language by new objects and methods for the representation
and manipulation of arbitrary symbolic expressions.
\vspace{1pc}
\end{abstract}

\maketitle

\section{Introduction}

\ginac is a \CC\ library that enables the user to perform algebraic operations within the \CC\
language.  It has been created to become the new algebra engine of the \xloops project
\cite{Bauer:2001ig} as a replacement for {\it Maple}.  Version $1.0$ released in the year 2001
finally provided all the necessary features for that intended purpose. Since then, \ginac has been
extended and improved continuously and is currently available in version 1.3.2.

A detailed discussion of the reasons why \ginac was started and how it was designed was given in
\cite{Bauer:2000cp} and doesn't need to be repeated here. But to point out the r\^ole of \ginac in
high energy physics calculations, we will shortly discuss the problem of software complexity.

When doing calculations in the context of perturbation theory, one has to accomplish a number of
tasks. The major ones are the following:
\begin{itemize}
\item Derivation of Feynman rules
\item Creating Feynman diagrams
\item Application of Feynman rules
\item Dirac algebra and calculating traces
\item Loop integration
\item Summation of diagrams
\item Renormalization
\item Phase space integration
\item Self-testing and comparisons
\end{itemize}
Each of these tasks poses certain mathematical or conceptual problems. Most can be addressed with
the help of computer programs. In certain cases the use of computers for automated calculation has
become essential. The list above can be categorized into a small number of algorithmic tasks:
\begin{itemize}
\item Combinatorics
\item Algebraic operations
\item Data management
\item Numerics
\item Input / Output
\end{itemize}
It is common practice to use specialized software tools for each of these tasks, like {\it Maple}
for algebra and {\it FORTRAN} for numerics. The software
components then have to be connected to each other by again different software tools to complete a
calculation.

Such automated calculations are confronted with a new kind of problem: software complexity. This
problem becomes more significant with the increasing scale of the calculations:
\begin{itemize}
\item Algorithms themselves are getting more complicated and involved.
\item The number of utilized algorithms is growing.
\item The number of different software technologies, i.e. programming languages or systems, in use
is rising.
\item More people are working on the same software project concurrently.
\end{itemize}

Thus topics like usability, verifiability, maintainability and extensibility become important.
Computer science deals with these issues since decades now and various solutions have been put forward.
Software engineering principles were invented and \CC\ has gained widespread acceptance as a means
to realize them. The question is, whether \CC\ can be used for high energy physics calculations in
that r\^ole as well.

\CC\ excels at combinatorics and input/output operations. It is also very good for doing numerics
and data management. Only one of the algorithmic tasks being listed above could not be done in \CC\
so far and that was algebraic operations. \ginac filled this gap. With \ginac it is now possible to
write software for physics computations completely in \CC\ adhering to software engineering
principles.

\section{Basic concepts}

\ginac is a software library that is used by writing \CC\ programs. It is not an algebra system with
an elaborate user interface. \ginac defines several new data types for \CC\ programs. A very important one
is the data type for indeterminate mathematical symbols.
Corresponding \CC\ variables can be defined like this
\begin{center}
\tt symbol x("x");
\end{center}
Here a \CC\ variable {\tt x} gets defined with the character string {\tt "x"} as its output name.

A very nice feature of \CC\ is the ability to re-define the built-in operators for the new data
types. For example, the $+$ operator can be used with symbols like
\begin{center}
\tt a + b
\end{center}
to form a symbolic expression. These expressions can be stored in another data type of \ginac
called {\tt ex}, which allows for the corresponding variables to contain arbitrary symbolic
expressions or numeric values. This can be seen in the following example.
\pagebreak

\begin{verbatim}
  #include <iostream>
  using namespace std;
  #include <ginac/ginac.h>
  using namespace GiNaC;

  int main()
  {
    symbol a("a"), b("b");
    ex myterm = sqrt(a*b) + 0.3*b; 
    cout << myterm.subs(b==2) << endl;
    return 0;
  }
\end{verbatim}

After some \CC\ commands to include the needed libraries, two symbols {\tt a} and {\tt b} and a
small formula are defined in the main program. The result of the substitution $b\!\rightarrow\!2$ is
printed on the screen.  The output is the expression $\sqrt{2a} + 0.6$. Apart from simple
operations like multiplication and summation one can see the usage of mathematical functions and the
easy mixing of symbolic and numeric terms. The integration into the \CC\ language is seamless.

\section{Feature overview}

We already introduced the two most important data types provided by \ginacnosp. Instead of 
listing the other data types available, it is more instructive to name the core
functionalities of \ginacnosp. 

Expressions in \ginac may not only consist of symbols but also of numbers. Numbers can be exact
including exact rational fractions, as well as floating point numbers with an arbitrary
numeric precision. Other objects that can be part of expressions are indexed variables like vectors or matrices.
Mathematical functions are another important example for allowed objects.

Expressions can be manipulated in various ways. Mathematical operations like symbolic derivation,
series expansion or symmetrization are available. Operations on polynomials like expanding and
collecting or square-free decomposition are also present. Pattern matching and algebraic
substitutions complete the picture.

Systems of linear equations can be solved. \ginac also knows about special algebras like the Clifford
algebra including the Dirac $\gamma^\mu$ matrices and the $SU(3)$ color algebra.

There are a lot of mathematical functions already defined in \ginac and new ones can easily be
added. All functions know their expected algebraic properties and can be evaluated numerically
for arbitrary complex arguments. In addition to the standard functions like $\sin$, $\cos$, Euler's
$\Gamma(x)$ function, $\sqrt{x}$, $\log$ and many more,
\ginac also provides polylogarithms of all kinds:
The classical polylogarithm $\mbox{Li}_n(x)$,
which includes the ubiquitous dilogarithm $\mbox{Li}_2(x)$, Nielsen's generalized polylogarithm
$\mbox{S}_{n,p}(x)$, the harmonic polylogarithm $\mbox{H}_{m_1,\ldots m_k}(x)$ \cite{RV00}, the
multiple zeta value $\zeta(m_1,\ldots,m_k)$ and the multiple polylogarithm
$\mbox{Li}_{m_1,\ldots,m_k}(x_1,\ldots,x_k)$ are available. 
Again, all these functions can be evaluated numerically for arbitrary complex arguments
\cite{Vollinga:2004sn}.

Finally, another important feature of \ginac has to be mentioned: \ginac is open-source
software licensed under the GPL. The code is well documented and a comprehensive manual is
available \cite{manual}.

\section{Summary}

We have given a short introduction to \ginacnosp. We highlighted the benefits of using \ginac in
complex software systems written for high energy physics calculations. Features of \ginac and a
programming example have been given. \ginac can be downloaded from {\tt http://www.ginac.de}.

\end{document}